\documentclass[sigconf]{acmart}

\usepackage{algorithmic}
\usepackage{graphicx}
\usepackage{textcomp}
\usepackage{hyperref}
\usepackage{tcolorbox}
\usepackage{booktabs}
\usepackage{multirow}
\usepackage{float}

\def\BibTeX{{\rm B\kern-.05em{\sc i\kern-.025em b}\kern-.08em
    T\kern-.1667em\lower.7ex\hbox{E}\kern-.125emX}}

\setcopyright{cc}
\setcctype{by}
\acmDOI{10.1145/3843282.3844431}
\acmYear{2026}
\copyrightyear{2026}
\acmISBN{979-8-4007-2985-0/2026/10}
\acmConference[AgenticDev '26]{Proceedings of the 1st International Workshop on Agentic AI for Next-Generation Software Development}{October 12--16, 2026}{Munich, Germany}
\acmBooktitle{Proceedings of the 1st International Workshop on Agentic AI for Next-Generation Software Development (AgenticDev '26), October 12--16, 2026, Munich, Germany}
\received{2026-07-15}
\received[accepted]{2026-08-20}

\begin{document}

\title{Agents as Architecture Evaluators: Automating ATAM and SAAM for Continuous Assessment}

\author{Adyansh Kakran}
\correspondingauthor
\orcid{0009-0002-1786-0069}
\affiliation{%
  \institution{Software Engineering Center Research, IIIT Hyderabad}
  \city{Hyderabad}
  \country{India}
}
\email{adyansh.kakran@research.iiit.ac.in}

\author{Sathvika Miryala}
\orcid{0000-0001-5206-6042}
\affiliation{%
  \institution{Software Engineering Center Research, IIIT Hyderabad}
  \city{Hyderabad}
  \country{India}
}
\email{miryala.sathvika@research.iiit.ac.in}

\author{Karthik Vaidyanathan}
\orcid{0000-0003-2317-6175}
\affiliation{%
  \institution{Software Engineering Center Research, IIIT Hyderabad}
  \city{Hyderabad}
  \country{India}
}
\email{karthik.vaidhyanathan@iiit.ac.in}

\renewcommand{\shortauthors}{Kakran, Miryala, and Vaidyanathan}


\begin{abstract}
The long-term success of a software system relies heavily on its underlying architecture, which dictates its performance with respect to quality attributes. Software Architecture Evaluation methods, such as the Architecture Tradeoff Analysis Method (ATAM) and the Software Architecture Analysis Method (SAAM), are widely recognized for their ability to mitigate risks and ensure reliable quality attributes. However, their widespread adoption remains limited due to the significant cost, time, and human expertise required for these processes. This paper investigates the potential of agentic AI to lower these barriers and facilitate architectural assessment.

To enable this preliminary study, we established a dataset by extracting human conducted evaluations from existing academic literature into a structured format. Using this dataset, we developed an agentic workflow to synthesize ATAM and SAAM evaluation reports based on these high level architectural descriptions. This approach simulates the reasoning process of human evaluators when provided with structured architectural data.

We validated the approach using automated metrics and human evaluation. Agentic workflows achieve strong relevance and completeness ratings of 4.5/5 and 3.9/5 respectively. The agents successfully automated the labor-intensive phases of scenario generation and risk identification, reducing the evaluation time from weeks to under \textbf{5 minutes} at a cost of less than \textbf{\$0.50} per run. The results confirm that agentic AI effectively lowers adoption barriers, enabling continuous, high-coverage architectural assessment.
\end{abstract}


\maketitle

\section{Introduction}
\label{sec:introduction}

Software architecture serves as the fundamental bridge between high-level requirements and technical implementation, providing the structural blueprint necessary to guide a system's development and evolution~\cite{bass2021software}. Evaluating Architecture is essential for ensuring quality attributes such as performance, security, and maintainability are addressed. Over the past two decades, established evaluation methods like ATAM, SAAM, and ALMA have been used to identify architectural risks and manage the complex tradeoffs between competing quality scenarios~\cite{sae_methods,sae_methods_2,sae_methods_3,lago_sae}.

Despite the rigor of these traditional methods, they often demand exhaustive, manual brainstorming sessions between architects and stakeholders. The recent surge in Generative AI and Large Language Models (LLMs) offers a promising path toward automation, with LLMs demonstrating significant potential in decision support, code generation, and documentation synthesis~\cite{esposito2025generative}. Capilla et al.~\cite{capilla_atam} developed an initial automation approach for architecture evaluation using single LLM interactions. However, given the inherently collaborative nature of architecture evaluations, we were motivated to design an agentic framework that better simulates multi-stakeholder collaboration. 

We propose a multi agent workflow as an early step to simulate a human evaluation team. Instead of processing raw enterprise artifacts, this system takes structured architectural summaries as input and synthesizes evaluation reports. We break the evaluation process into specific roles and assign them to autonomous agents. These agents plan, review, and refine tasks to approximate the structured interactions described by Kazman et al. \cite{kazman_atam,kazman_saam}.

A significant hurdle in validating software architecture related systems is the lack of structured datasets. Unlike code generation tasks, architectural knowledge is typically locked in unstructured PDF reports. To evaluate this workflow, we curated a dataset by extracting human conducted evaluations for SAAM and ATAM from academic literature into a structured JSON format. The \textbf{main contributions} of this paper include: 
\begin{itemize} 
    \item A dataset of 18 ATAM and 21 SAAM structured evaluations.
    \item An agentic workflow to simulate the architectural evaluations.
\end{itemize}

The remainder of this paper is organized as follows: Section \ref{sec:background} provides the necessary background while Section \ref{sec:related_work} presents the related work. Section \ref{sec:dataset} elaborates the construction of dataset. Section \ref{sec:methodology} details the design of the agentic workflow. Section \ref{sec:results} presents results, followed by a discussions in Section \ref{sec:discussion} and threats to validity in Section \ref{sec:threats_to_validity}. Section \ref{sec:conclusion} concludes the study~\footnote{The data and code for this study is available at \url{https://zenodo.org/records/21374126}}.

\section{Background}
\label{sec:background}

\subsection{Software Architecture Evaluation}
Software architecture serves as the blueprint for a system. Software Architecture Evaluation (SAE) methods are designed to identify risks early by evaluating the extent to which a proposed architecture meets its quality attribute requirements.

Despite the benefits of early evaluation, widely accepted methods such as SAAM and ATAM face adoption challenges~\cite{sae_adoption_barriers}. Their reliance on extensive stakeholder participation over multiple days makes them resource-intensive, which often leads to their omission from the Software Development Life Cycle (SDLC), particularly in constrained environments.

We focus on two of the most popular~\cite{sae_methods,sae_methods_2} software architecture evaluation techniques:

\subsubsection{Software Architecture Analysis Method (SAAM)}
This method of evaluation focuses primarily on the quality attribute of modifiability~\cite{kazman_saam}. It employs a scenario-based approach where stakeholders enumerate potential future uses or changes to the system.  The scenarios are classified as \textit{direct} (supported by the current architecture) or \textit{indirect} (requiring structural modification). The end result of the evaluation is a modifiability assessment and general architectural strengths and weaknesses of the proposed architecture.

\subsubsection{Architecture Tradeoff Analysis Method (ATAM)}
Building upon the foundations of SAAM, ATAM expands the scope to evaluate multiple competing quality attributes~\cite{kazman_atam}. ATAM is distinct in its focus on the interaction between these attributes. The method utilizes a Utility Tree to prioritize business drivers and map them to specific architectural decisions. The primary outputs of an ATAM evaluation are architectural risks, non risks, sensitivity points and tradeoff points.

\subsection{LLMs and Agentic AI in Software Engineering}
The application of LLMs in software engineering has expanded beyond code generation to include high-level architectural tasks. Recent research has demonstrated the utility of LLMs in managing architectural knowledge, such as recovering architectural information from legacy codebases~\cite{pan2026archagentscalablelegacysoftware}. While singular model interactions are effective for isolated queries, Software Architecture Evaluation requires maintaining extensive system context and reasoning.

To support these complex, multi-step processes, the field has moved toward Agentic AI. Unlike standard interaction models, agentic workflows utilize LLMs as reasoning engines within a framework that supports autonomous planning, memory management, and tool usage~\cite{agents_in_se}. In software engineering, multi-agent systems have the advantage of role-based collaboration, which is particularly pertinent to architecture evaluation, as it mirrors the stakeholder interactions and varied perspectives required by methods like ATAM and SAAM.

\section{Related Work}
\label{sec:related_work}

Although methods such as ATAM and SAAM effectively identify risks, their industrial adoption remains limited due to high resource demands~\cite{sae_adoption_barriers}. Sahlabadi et al. highlight that the requirement for synchronous stakeholder participation often renders these methods incompatible with modern agile processes~\cite{lightweight_eval}. Consequently, organizations frequently omit formal evaluation, which leads to the silent accumulation of architectural debt~\cite{sae_adoption_barriers}.

To reduce the costs, earlier automation attempts, such as those by Alzamil et al., utilized quantitative metrics like coupling and cohesion~\cite{coupling_cohesion}. However, rule-based approaches fail to capture the semantic nuances necessary to evaluate trade-offs against business drivers~\cite{sa_in_practice}. As shown by Scalabrino et al., while quantitative metrics effectively detect structural violations, they cannot fully replace qualitative judgment in assessing quality attributes or understanding code intent~\cite{code_understanding}.

Recent advances in LLMs have introduced new avenues for semantic processing in software engineering~\cite{esposito2025generative}. In the realm of architectural knowledge, Dhar et al. utilized LLMs to 
generate Architectural Decision Records~\cite{dhar}, while Santos et al. explored LLM augmentation across the full development lifecycle~\cite{llm_sdlc}. For architecture evaluations in particular, Capilla et al. analyzed the performance of Microsoft Copilot in conducting ATAM evaluations~\cite{capilla_atam}.

To overcome the reasoning limitations of single models, recent research has pivoted toward multi-agent systems that simulate human collaboration~\cite{chatdev}. Frameworks like MetaGPT illustrate that assigning distinct roles significantly enhances performance on complex tasks~\cite{metagpt}. While Wang et al. underscore the significant potential for agentic systems in software engineering~\cite{agents_in_se}, existing workflows predominantly focus on implementation and code generation rather than high-level architectural assessment. To the best of our knowledge, no prior work has leveraged role-based agentic workflows to automate the multi-stakeholder dynamics of ATAM and SAAM.

\section{Dataset creation}
\label{sec:dataset}

\begin{figure}[htbp]
\centering
\includegraphics{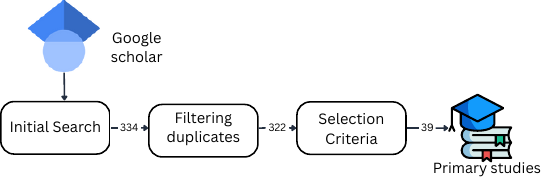}
\caption{Workflow for Selection Process}
\label{search_process}
\end{figure}
\begin{figure*}[t]
  \centering
  \includegraphics[width=\textwidth]{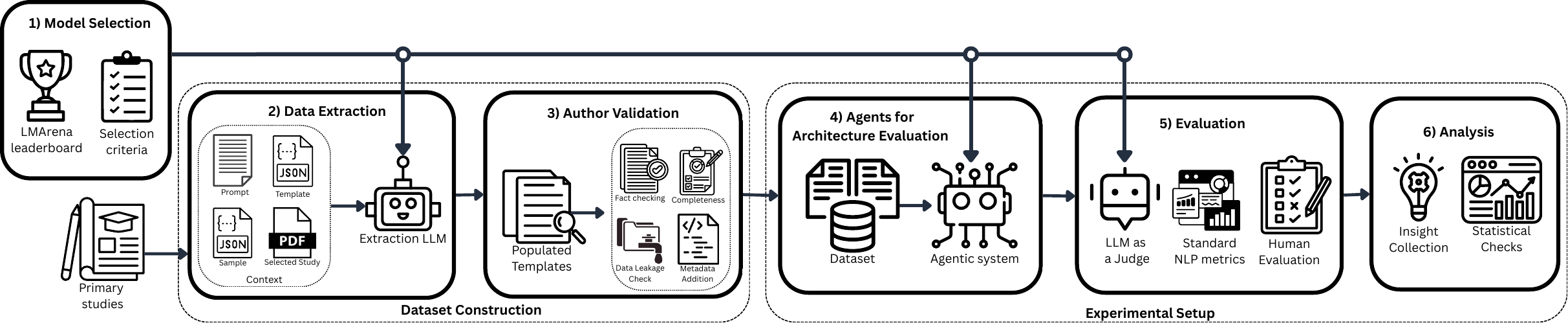}
  \caption{Experimental Design}
  \label{study_design}
\end{figure*}

To evaluate our approach, we required architectural descriptions and corresponding human conducted evaluations, which we can use as ground truth. A significant challenge in this domain is that public software architecture evaluation studies are predominantly available as unstructured PDFs. To the best of our knowledge, there is currently no publicly available dataset that maps architectural descriptions to their corresponding evaluations.

To address the gap and validate the efficacy of our agentic workflow, we curated a dataset by extracting structured evaluation reports from academic literature. This section details the dataset curation pipeline, from case study selection to human-verification of the data.

\subsection{Primary Study Selection}

To ensure our dataset represents a diverse set of architectural styles, and complexity levels, we employed a systematic selection strategy. The selection workflow is visualized in Figure \ref{search_process}.

\subsubsection{Initial Search}

To achieve comprehensive coverage of both academic literature and industrial technical reports, we utilized \href{scholar.google.com}{Google Scholar} as our primary index. We used a query string for selection of case studies from the platform for both evaluation methods. The query for SAAM was ``\textit{intext: SAAM OR intext: Software Architecture Evaluation Method AND intitle: case study}``, with an analogous query constructed for ATAM.

These search terms were designed to prioritize precision over recall, ensuring a dataset of sufficient quality to validate our proposed methodology. The search strategy explicitly targeted papers documenting practical applications of ATAM or SAAM, thereby filtering out theoretical frameworks or methodological proposals lacking an execution of the evaluation. The initial search yielded a total of $334$ candidate case studies.

\subsubsection{Filtering Duplicates}

Since Google Scholar aggregates different versions of the same publication (e.g., preprints, conference proceedings, and extended journal articles), the dataset contained some redundant entries. We applied a manual deduplication process, keeping the latest archival version of each eligible paper. This step reduced the candidate pool to 322 unique studies.

\subsubsection{Selection Criteria}

Each datapoint in the dataset should contain sufficient information for a general software architecture evaluation report. To this end, the case studies were screened against strict eligibility criteria. 

\textbf{Inclusion Criteria (IC):}
\begin{itemize}
    \item \textit{IC1:} The study explicitly applies ATAM or SAAM as the primary evaluations.
    \item \textit{IC2:} The study contains a concrete, real-world or academic case study (not just a theoretical proposal).
    \item \textit{IC3:} The study provides a sufficient description of the software architecture to allow for re-evaluation.
    \item \textit{IC4:} The study explicitly lists the evaluation results that serve as ground truth.
    \item \textit{IC5:} The full text is available in English.
\end{itemize}

\textbf{Exclusion Criteria (EC):}
\begin{itemize}
    \item \textit{EC1:} Systematic Literature Reviews (SLRs), surveys, or meta-analyses were excluded as they lack evaluation data.
    \item \textit{EC2:} Studies that mention performing an ATAM but do not publish specific findings (e.g., stating "we found 5 risks" without detailing them) were excluded.
    \item \textit{EC3:} Studies performing  variants of SAAM or ATAM that deviate significantly from the standard methodology were excluded.
\end{itemize}

Following this screening, the final dataset comprised 39 case studies, with 18 ATAM and 21 SAAM evaluations, suitable for extraction.

\subsection{Template Creation}
\label{subsec:template_creation}

A major challenge in formalizing these evaluations is the heterogeneity of natural language. To convert unstructured academic articles into a structured and standardized format, we designed a JSON schema for both ATAM and SAAM. This standardization decouples the evaluation logic from the different writing styles of different authors. Due to space limitations, we present a summary of the templates in the text boxes. The complete JSON schemas for both ATAM and SAAM are available in our replication package repository. These schemas detail the strict data types and required fields that guide the agents during the data extraction and evaluation steps.

\begin{tcolorbox}[
  colback=black!5!white,
  colframe=black!50!black,
  colbacktitle=black!75!black,
  title=SAAM Evaluation Template,
  fonttitle=\bfseries
]

\textbf{Metadata:}
Paper identification, Authors, venue, year

\textbf{Evaluation Inputs}
\begin{itemize}
  \item \textit{System Context}: domain, purpose, maturity
  \item \textit{Evaluation Setup}: goal, type, architectures
  \item \textit{Architecture Description}: components, styles
  \item \textit{Scenario Elicitation}: sources, types
  \item \textit{Evaluation Focus}: quality attributes
\end{itemize}

\textbf{Ground Truth Outputs}
\begin{itemize}
  \item Scenarios and classifications
  \item Scenario-based change analysis
  \item Inter-scenario interactions
  \item Overall assessment and recommendations
\end{itemize}

\end{tcolorbox}

Another motivation to create this template was the separatation of data into two logically distinct sections to prevent data leakage during experimentation:
\begin{enumerate}
    \item \textbf{Evaluation Inputs:} The set of inputs only the information available to an evaluator \textit{before} the analysis begins. It includes business drivers, architectural information, and stakeholder constraints.
    \item \textbf{Ground Truth Outputs:} It captures the findings of the human experts, including identified risks, prioritized scenarios and recommendations.
\end{enumerate}

\subsection{Data Extraction}

Since their inception, large language models (LLMs) have demonstrated strong capability in transforming unstructured textual data into structured representations. There have been multiple studies that document and validate this use case for LLMs \cite{llm_data_extraction, llm_data_extraction_2}. This automation enables the generation of candidate templates with significantly reduced manual effort.

\begin{tcolorbox}[
  colback=black!5!white,
  colframe=black!50!black,
  colbacktitle=black!75!black,
  title=ATAM Evaluation Template,
  fonttitle=\bfseries
]

\textbf{Metadata}
\begin{itemize}
  \item System identification and date
  \item Evaluator organization
  \item Phase 1 \& 2 attendees (roles/titles)
\end{itemize}

\textbf{Evaluation Inputs}
\begin{itemize}
  \item \textit{Business Context}: goals, constraints, drivers
  \item \textit{Phase 1 Discovery}: documentation, questions
  \item \textit{Phase 2 Context}: hardware, runtime constraints
  \item \textit{Quantitative Parameters}: quantifiable attributes
  \item \textit{Investigative Prompts}: style probes, raw scenarios
\end{itemize}

\textbf{Ground Truth Outputs}
\begin{itemize}
  \item \textit{Utility Tree}: attributes, refinements, priority tuples
  \item \textit{Architectural Approaches}: patterns and rationale
  \item \textit{Analysis Records}: decisions, reasoning
  \item \textit{Findings}: risks/non-risks, tradeoffs
  \item \textit{Final Scenarios}: prioritized scenarios
\end{itemize}

\end{tcolorbox}

We employed a \textit{Context-Aware Few-Shot Prompting} strategy, to enable in-context learning~\cite{kwak-etal-2024-classify}. The prompt structure provided to the LLM included:

\begin{enumerate}
    \item \textbf{Process Definition:} Description of the respective Software Architecture Evaluation Process.
    \item \textbf{Role Definition:} Establishing the persona of a ``Software Architecture Researcher.'' and giving the details of the task at hand.
    \item \textbf{Template Definition:} The strict JSON structure required for the output.
    \item \textbf{One-Shot Example:} A manually extracted case study to demonstrate the desired level of granularity (e.g., how to summarize a `Risk' without losing technical nuance).
\end{enumerate}

The LLM processed the raw text and mapped the unstructured PDFs into our predefined JSON fields, effectively standardizing terminology across different studies.

\subsection{Author Validation}

While LLMs are great at syntactic structuring, they are prone to hallucinations or missing out details~\cite{llm_data_extraction_challenges}. To ensure the integrity of our dataset, we implemented a human-in-the-loop verification process. Two of the authors performed verified every generated JSON file against the original source text.

This validation process focused on three key dimensions of data quality:

\begin{itemize}
    \item \textbf{Factuality Check:} We verified that every technical claim, risk, or scenario present in the JSON was supported by the source text. Any inferred or extrapolated details were removed to ensure the dataset reflects only the original authors' intent.
    \item \textbf{Completeness:} We cross-referenced the extraction against the paper's diagrams and architecture description to ensure no critical architectural constraints or non-functional requirements were missed. This was crucial for the ``Evaluation Inputs'', as missing context would unfairly penalize the agent.
    \item \textbf{Data Leakage:} We checked for data leakage, ensuring that information belonging to the ``Ground Truth Outputs'' (e.g., a specific risk found during analysis) was not accidentally categorized into the ``Evaluation Inputs''. 
\end{itemize}

\section{Methodology}
\label{sec:methodology}

\begin{figure}[htbp]
    \centering
    \includegraphics[width=0.5\textwidth]{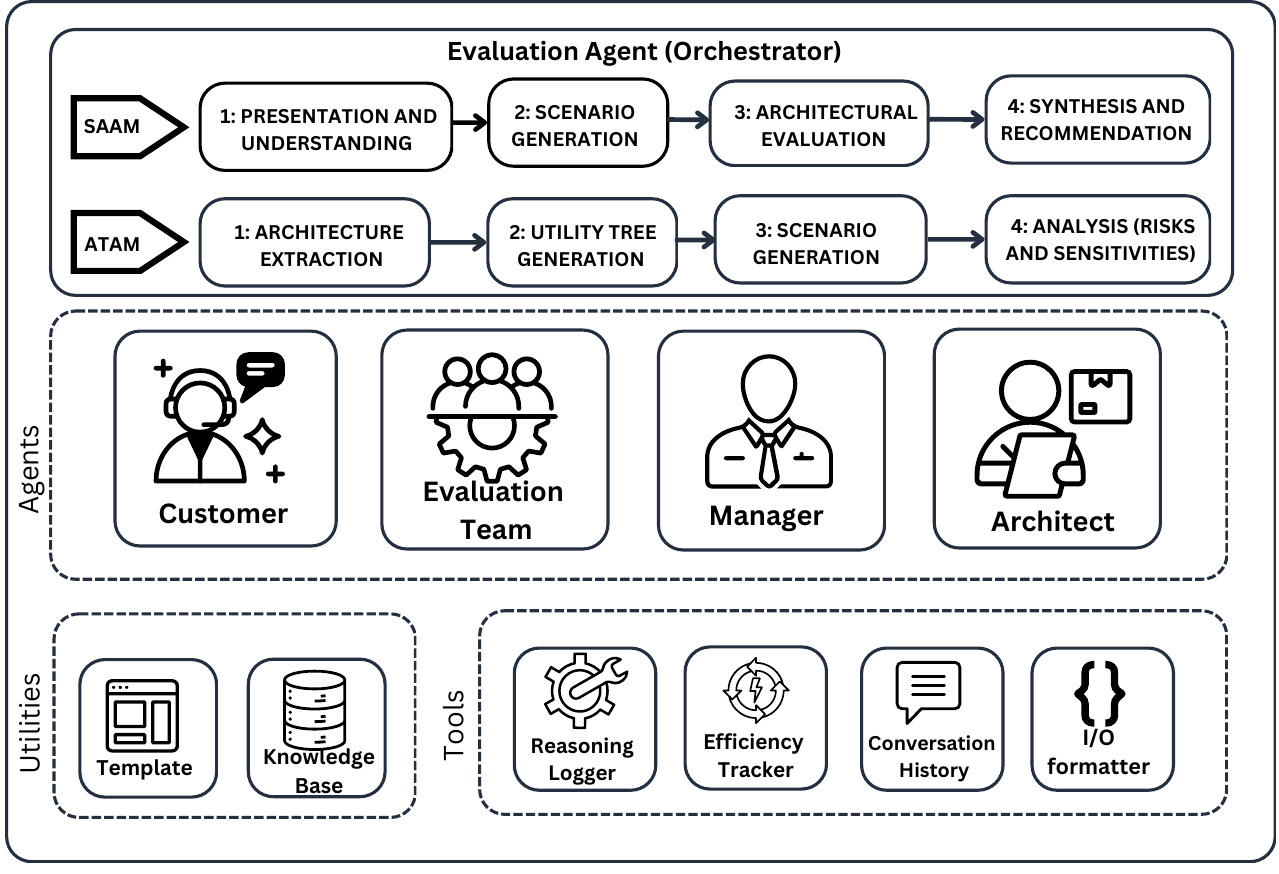}
    \caption{Main Architecture for both Workflows}
    \label{agent_arch}
\end{figure}

Figure~\ref{agent_arch} illustrates the architecture of agentic framework. As shown in the figure, the framework is designed to autonomously execute the ATAM and SAAM through coordinated role-based agents. Further, we detail the orchestration logic, the phase-by-phase workflow, and the persona engineering strategy that collectively enable the automated evaluations.

\subsection{Agents for Architecture Evaluation}

The workflow is governed by a deterministic \textit{Orchestrator} script that manages the evaluation state and enforces task sequencing. This central controller directs four specialized LLM agents. The agents communicate strictly through a shared JSON blackboard rather than a direct chat interface. When the Orchestrator assigns a task, the designated agent reads the current state of the blackboard, processes its specific role, and writes updates back to the file. To manage context limits effectively during these interactions, each agent maintains its own interaction history using a sliding window approach. This step by step process allows the system to simulate a team working on a single evaluation document. The \textit{Customer} agent provides architectural information and clarifies domain constraints. The \textit{Manager} acts as a facilitator, translating vague stakeholder inputs into formal requirements. The \textit{Architect} performs the core technical analysis and formalizes system descriptions. Finally, the \textit{Evaluation Team} synthesizes these outputs to identify risks, trade-offs and conclude the whole analysis with the final results. The agents are designed to be generalisable, but are pertinent to the SAAM and ATAM evaluation techniques.

The interaction between agents is strictly controlled by the Orchestrator to prevent endless conversation loops. When an agent receives a turn, it is provided with its specific system prompt, the current state of the JSON blackboard, and a short memory of the last three actions. We use a structured debate approach to refine the outputs. For example, to gather stakeholder requirements, the Manager agent is prompted to read the business drivers from the blackboard and propose a scenario. The Architect agent is then prompted to read this proposed scenario and check it against the recorded system constraints. If the Architect finds a technical conflict, it writes a rejection note to the blackboard. The Orchestrator then assigns the task back to the Manager to revise the scenario. This review step ensures that outputs are checked by a different role before they are finalized in the template.

\subsubsection{\textbf{The SAAM Workflow}}
\begin{figure}[htbp]
    \centering
    \includegraphics[width=0.5\textwidth]{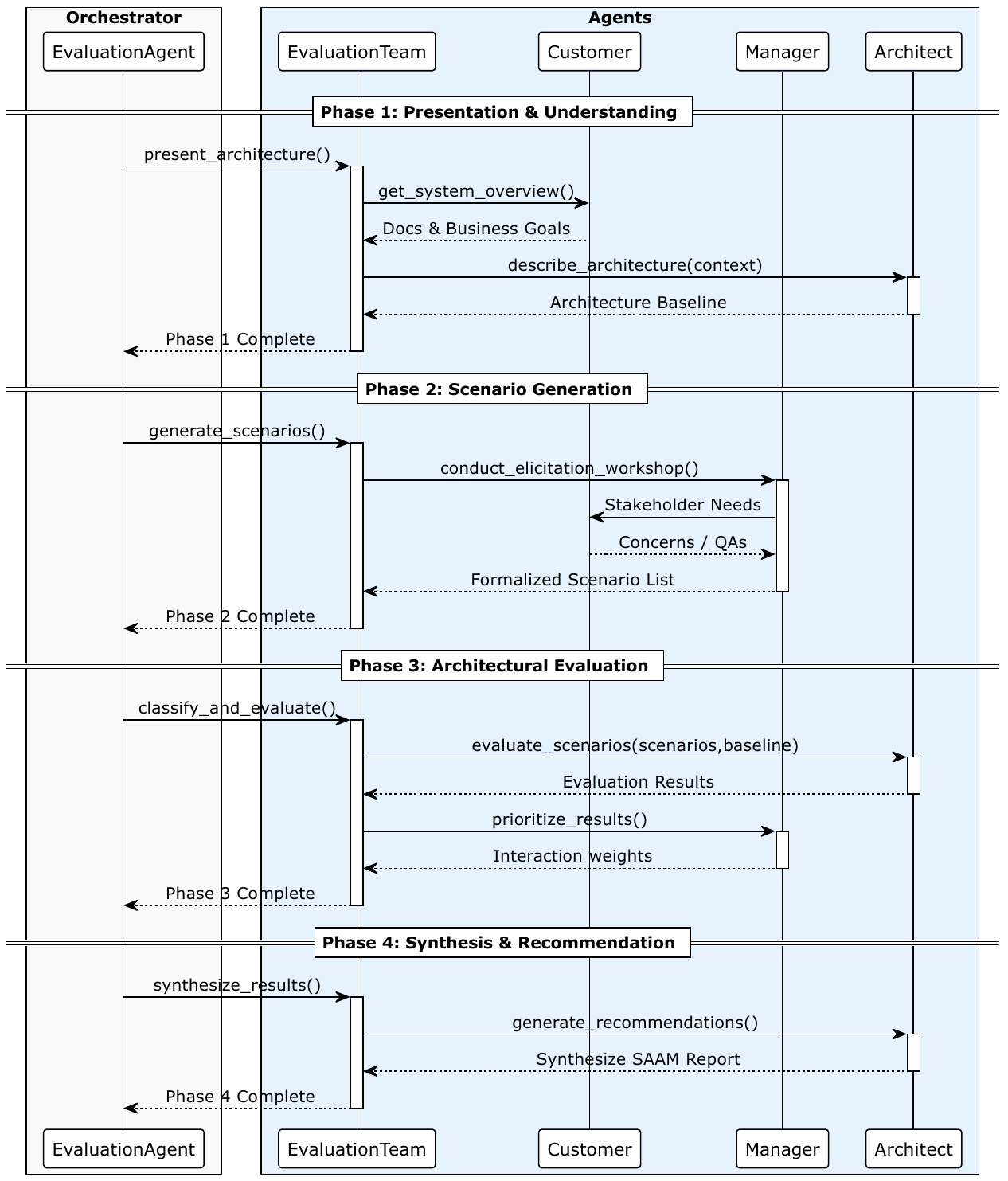}
    \caption{Sequence Diagram for SAAM}
    \label{saam_sequence}
\end{figure}

The workflow is divided into multiple phases according to major tasks as directed in the original SAAM study by Kazman et al. \cite{kazman_saam}. The flow is visualized in Figure \ref{saam_sequence}.

\paragraph{\textbf{Phase 1: Architecture Presentation and Understanding}} The process begins with the \textit{Architect} establishing a shared understanding of the system. The \textit{Architect} compares the provided documentation against the required fields in the JSON template. If a required field is missing or unclear, the \textit{Architect} formulates a question. For example, if the text mentions a database but lacks performance details, the \textit{Architect} might ask, "What is the expected transaction volume for the database component?" and writes this query to the blackboard. The \textit{Orchestrator} then prompts the \textit{Customer} to check the blackboard. The \textit{Customer} reads the question, reviews the raw domain descriptions, and posts the answer back to the blackboard. Once the details are clear, the \textit{Architect} reads the answer and updates the strict JSON schema. This defines components, connectors, and their associated quality attributes for the whole workflow to use internally.

\paragraph{\textbf{Phase 2: Scenario Elicitation}} The \textit{Manager} initiates a dialogue with the \textit{Customer} to gather stakeholder needs. Unlike a simple transcription, the \textit{Manager} uses the architecture description from Phase 1 to filter these requests. It distinguishes between simple feature requests and true architectural drivers. These are then formatted into standard SAAM scenarios, comprising a stimulus, response, and the primary quality attribute it affects.

\paragraph{\textbf{Phase 3: Architecture Evaluation}} The \textit{Architect} assesses the impact of each scenario on the formalized architecture. It classifies scenarios as direct (supported) or indirect (requires modification). For indirect scenarios, the agent estimates the change effort and identifies the specific components involved. Following this, an interaction analysis to detect structural clustering, identifying components that are affected by multiple scenarios. This analysis identifies conflicts, where scenarios impose contradictory requirements and coupling. Finally the \textit{Manager} then prioritizes these interactions against business goals, ensuring they address high value concerns and not small implementation details.

\paragraph{\textbf{Phase 4: Synthesis and Recommendations}} In the final phase, the \textit{Architect} aggregates the interaction data and scenario classifications. It groups direct scenarios to highlight system strengths and indirect scenarios to pinpoint weaknesses. The agent calculates a qualitative modifiability metric based on the ratio of direct to indirect scenarios. Finally, the \textit{Evaluation Team} generates a prioritized list of recommendations, ranking architectural changes by their estimated effort and impact on business goals.
\\
\subsubsection{\textbf{The ATAM Workflow}}

\begin{figure}[htbp]
    \centering
    \includegraphics[width=0.5\textwidth]{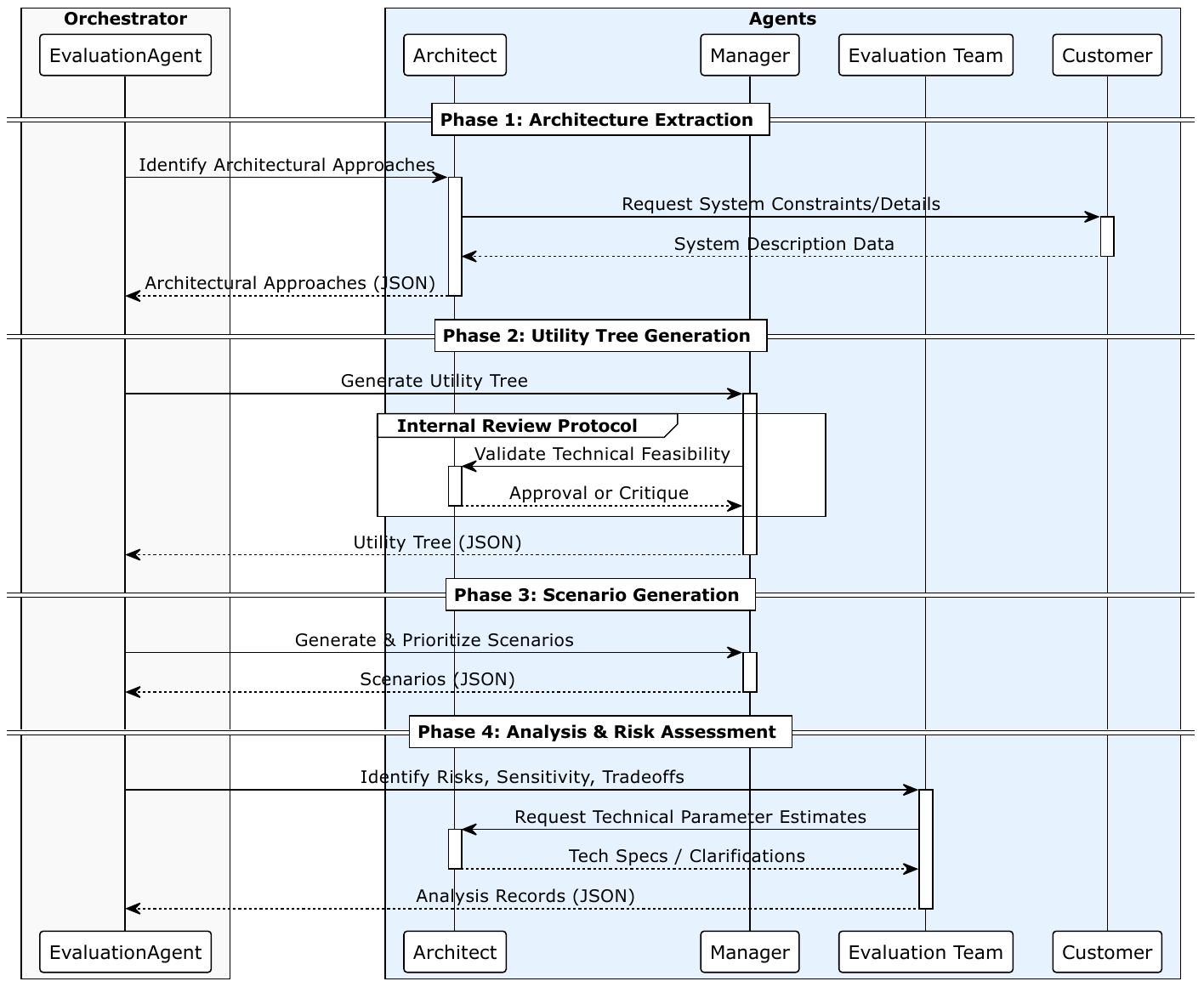}
    \caption{Sequence Diagram for ATAM}
    \label{atam_sequence}
\end{figure}

To automate an ATAM, we split the standard process into four distinct phases, each managed by a specific agent as shown in Figure \ref{atam_sequence}. This division of responsibilities was taken from the original ATAM paper by Kazman et al. \cite{kazman_atam}, making each major task in the process a phase.

The workflow proceeds as follows:

\paragraph{\textbf{Phase 1: Architecture Extraction}} The \textit{Architect's} first job is to synthesize raw documentation into a structured format. It actively queries the \textit{Customer} to clarify system constraints and details before converting it into a structured section called ``Architectural Approaches''. This output also forces the agent to map these styles to specific quality attributes, providing a rationale for why a certain approach was chosen.

\paragraph{\textbf{Phase 2: Utility Tree Generation}}
Once the architecture is documented, the \textit{Manager} takes over to build the Utility Tree. This step translates high level business goals into quality attributes and corresponding scenarios. To prevent the generation of unrealistic scenarios, we built an Internal Review Protocol. The \textit{Manager} drafts a scenario and posts it to the blackboard for the \textit{Architect} to validate. For instance, if the \textit{Manager} proposes a scenario requiring zero downtime during major updates, the \textit{Architect} reads the proposal from the blackboard and reviews the current architecture constraints to see if that is technically possible. If the \textit{Architect} approves and updates the status on the blackboard, the scenario is finalized. The final output is a hierarchical tree where each quality attribute and corresponding scenario is ranked by a priority tuple. This captures the importance to the stakeholder and the difficulty of handling it.

\paragraph{\textbf{Phase 3: Scenario Elicitation}}
With the priorities set, the \textit{Manager} moves to broader scenario generation. While Phase 2 focuses on high-priority business items, Phase 3 is about coverage. The agent brainstorms a mix of use-case, growth, and scenarios to stress-test the system boundaries. These are saved into the scenarios schema, which pairs the full textual description of the event with its primary quality attribute.

\paragraph{\textbf{Phase 4: Analysis \& Risk Assessment}}
The final phase is handled by the \textit{Evaluation Team}. This agent looks at the data holistically, identifying all possible avenues of architectural debt. It may request technical parameter estimates from the \textit{Architect} to clarify uncertainties. The final output is recorded in the JSON schema, where findings are categorized into \textit{risks}, \textit{non\_risks}, \textit{sensitivity\_points} and \textit{tradeoff\_points}.

\subsection{Reasoning and Metric Collection}

To understand not just the quality of the output but also the cost, we built two logging mechanisms into the framework:

\begin{itemize}
    \item \textbf{Efficiency Tracker:} Since agentic workflows are prone to trigger a cascade of API calls due to repeated back-and-forth communication between the agents, this tracker logs the token usage and latency for every interaction. This gave us the data needed to estimate the actual economic cost of running an automated ATAM versus a human one.

    \item \textbf{Reasoning Logger:} We instructed every agent to prepend a reasoning block to their JSON outputs. This acts as a kind of internal monologue, explaining why a risk was flagged or a scenario was prioritized. This reasoning trail was helpful during validation, helping us distinguish between sound logical conclusions and occasional hallucinations.
\end{itemize}

\subsection{Persona Design and Prompting Techniques}

A major benefit of using multi agent systems is the ability to use multiple different roles for each task. We used Persona based prompting to enforce these boundaries \cite{persona_multi_agent}. For example, the \textit{Architect} is prompted to be defensive and technical, while the \textit{Evaluation Team} is told to be skeptical and critical. This manufactured conflict prevents the agents from simply agreeing with each other, forcing the kind of debate that happens in real architectural evaluations.

Finally, enforcing our JSON schema was a significant challenge. To handle this, we used a Self-Correcting Prompt Loop. If an agent generated an invalid JSON, the error was caught and fed back to the Orchestrator, which then re-prompted the agent to fix the syntax. This loop ensured that the final outputs were always structured correctly.

\section{Results}
\label{sec:results}

The experiments are designed to address the following research questions, which we use to evaluate the data extraction process and agentic workflow.

\noindent \textbf{RQ1:} \textit{How accurately can LLMs extract unstructured architectural data and populate formal evaluation templates?}

\noindent \textbf{RQ2:} \textit{To what extent can agentic workflows replicate human-led software architecture evaluations?}
\begin{itemize}
    \item \textbf{RQ2.1:} \textit{How do agent-generated evaluations compare to human baselines across automated metrics?}
    \item \textbf{RQ2.2} \textit{How does the quality of agentic evaluations differ from human teams in terms of quality, as judged by humans?}
\end{itemize}

\noindent \textbf{RQ3}: \textit{What are the cost and sustainability implications of automating software architecture evaluation?}

\subsection{Experimental Setup}

To assess the capabilities of our agentic framework, we executed the evaluation pipeline against all the verified case studies in our dataset. To ensure reproducibility and minimize variance during the generation of the evaluation reports, we set the temperature to $0.2$ for all model inferences.

\subsubsection*{\textbf{Model Selection}}

Since our study utilizes LLMs in multiple distinct use cases, model selection was critical. We referenced the \textit{LMSYS Chatbot Arena Leaderboard}~\footnote{https://arena.ai/leaderboard/text} to identify state-of-the-art models. Table \ref{tab:models_selected} summarizes the models selected for different tasks in this study. We selected \textit{Gemini 3 Flash} for high-volume tasks (Data Extraction, Judging) due to its cost-efficiency. For the core Agentic Workflow, we evaluated three distinct model families: \textit{Gemini 3 Flash}, \textit{Claude 4.5 Sonnet}, and \textit{DeepSeek v3.2}. This selection allows us to benchmark proprietary efficiency models against open-weights alternatives, providing a representative snapshot of the state-of-the-art at the time our experiments were conducted.

\begin{table}[H]
\caption{Models Selected}
\label{tab:models_selected}
\centering
\renewcommand{\arraystretch}{1.2}

\begin{tabular}{llcc}
\toprule
\textbf{Task} & \textbf{Model} & \textbf{Context} & \textbf{Open Source} \\
\midrule

Data Extraction 
& Gemini 3 Flash 
& 1M 
& No \\

\midrule

\multirow{3}{*}{Agentic Workflow} 
& Claude 4.5 Sonnet 
& 1M 
& No \\
& Gemini 3 Flash 
& 1M 
& No \\
& Deepseek v3.2 
& 128K 
& Yes \\

\midrule

LLM-as-a-judge 
& Gemini 3 Flash 
& 1M 
& No \\

\bottomrule
\end{tabular}

\end{table}

\subsection{Evaluation Metrics}

Evaluating the output of a subjective process like software architecture evaluation is challenging. To address this, we employed a multi-faceted evaluation strategy. We utilized four standard NLP metrics-\textbf{ROUGE-L}~\cite{ROUGE}, \textbf{BLEU}~\cite{BLEU}, and \textbf{METEOR}~\cite{METEOR}, and \textbf{BERTScore}~\cite{bertscore}. These metrics provide a quantitative baseline for lexical and semantic similarity between the agent-generated reports and the ground-truth human evaluations.

To bridge the gap between statistical similarity and human utility, we conducted both an \textbf{LLM-as-a-Judge}~\cite{li2024llmsasjudgescomprehensivesurveyllmbased} evaluation and a \textbf{Human Evaluation}, performed by two of the authors. For both evaluations, we established a rubric based on a 5-point Likert scale across four dimensions:

\begin{itemize}
    \item \textbf{Relevance:} The degree to which the generated risks and scenarios pertain specifically to the architecture under review.
    \item \textbf{Coherence:} The logical flow of the evaluation, ensuring that identified risks are linked to the final architectural decisions.
    \item \textbf{Completeness:} The extent to which the agent identified critical risks and sensitivity points that were flagged in the ground-truth human expert reports.
    \item \textbf{Conciseness:} The information density of the report, penalizing verbosity or repetitive filler text that dilutes technical insights.
\end{itemize}

\begin{table*}[t]
\centering
\caption{Evaluation Metrics for Agent Generated Software Architecture Evaluations}
\label{tab:results_master}
\resizebox{\textwidth}{!}{%
\begin{tabular}{llcccccccc}
\toprule
 &  & \multicolumn{4}{c}{\textbf{LLM-as-a-Judge}} & \multicolumn{4}{c}{\textbf{Human Evaluation}} \\
\cmidrule(lr){3-6} \cmidrule(lr){7-10}
\textbf{Method} & \textbf{Model} 
& \textbf{Relevance} & \textbf{Coherence} & \textbf{Completeness} & \textbf{Conciseness} 
& \textbf{Relevance} & \textbf{Coherence} & \textbf{Completeness} & \textbf{Conciseness} \\
\midrule

\multirow{3}{*}{\textbf{SAAM}} 
& Gemini 3 Flash 
& 3.738 & 4.429 & 3.179 & \textbf{4.191} 
& \textbf{4.375} & 4.292 & 3.708 & \textbf{4.542} \\

& DeepSeek v3.2 
& 3.886 & 4.648 & 3.239 & 4.102 
& 4.292 & \textbf{4.375} & 3.750 & 4.292 \\

& Claude 4.5 Sonnet 
& \textbf{4.466} & \textbf{4.83} & \textbf{4.227} & 3.886 
& 3.917 & 4.292 & \textbf{3.875} & 2.958 \\

\midrule

\multirow{3}{*}{\textbf{ATAM}} 
& Gemini 3 Flash 
& 4.444 & \textbf{4.931} & 3.583 & \textbf{4.875} 
& \textbf{4.545} & \textbf{4.227} & 3.455 & \textbf{4.364} \\

& DeepSeek v3.2 
& 4.25 & 4.708 & 3.306 & 4.431 
& 4.136 & 3.864 & 3.182 & 3.864 \\

& Claude 4.5 Sonnet 
& \textbf{4.639} & 4.722 & \textbf{4.25} & 4.139 
& 4.182 & 4.136 & \textbf{3.636} & 3.182 \\

\bottomrule
\end{tabular}%
}
\end{table*}

In the following subsections, we analyze these results in detail, addressing each research question sequentially to validate both the quality of our ground-truth dataset and the efficacy of the agentic evaluations.

\subsection*{\textbf{RQ1} Data Extraction Performance}

The performance of the agentic workflow is predicated on the quality of the input data. Our extraction pipeline processed a total of 39 primary studies (18 ATAM and 21 SAAM). The results indicate that modern LLMs are highly suited for syntactic structuring, the extraction process yielded zero JSON schema violations, confirming that the model could strictly adhere to the complex template definitions provided in Section \ref{subsec:template_creation}.

However, information retrieval required human intervention. During the validation phase, we identified two primary error categories:
\begin{enumerate}
    \item \textbf{Missing key information:} In 7 of the 39 cases ($\approx$ 18\%), the LLM failed to capture niche scenarios that were present in the given reference texts, and the authors had to manually add them to the populated template. 
    \item \textbf{Hallucination:} In 2 cases ($\approx$ 6\%), the model fabricated minor details to fill fields where the original text was ambiguous, despite instructions in the prompt to mark such fields as ``Unknown.''
\end{enumerate}

The diverse dataset resulting from this human-verified extraction serves as the foundation for the subsequent performance analysis.

\subsection*{\textbf{RQ2.1} Automated Metrics}

\begin{figure}[!ht]
    \centering
    \includegraphics[width=0.5\textwidth]{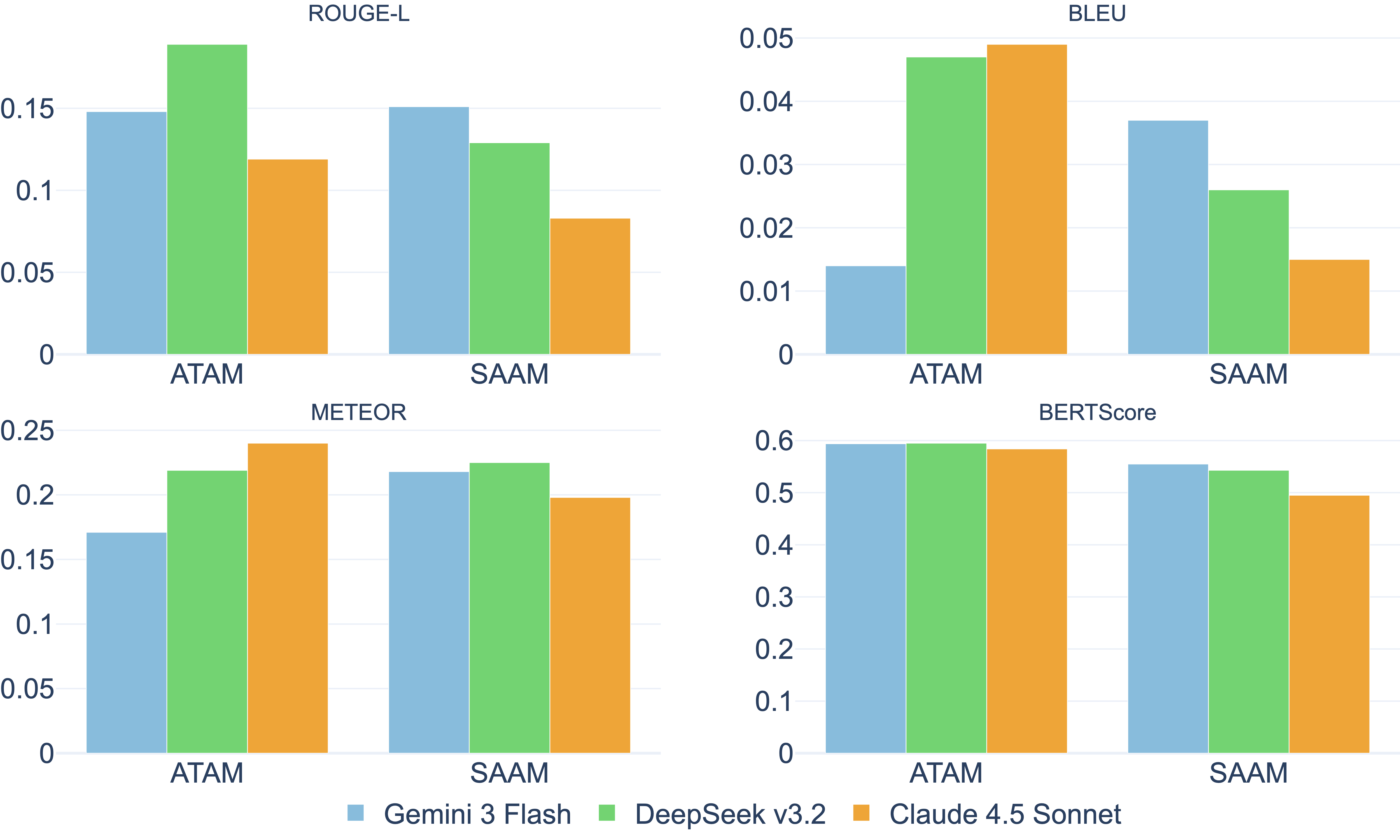}
    \caption{Lexical and Semantic Evaluation Metrics}
    \label{lexical}
\end{figure}

Figure \ref{lexical} shows the performance of our agentic system on the lexical and semantic based metrics. Table \ref{tab:results_master} summarizes the performance for LLM-as-a-judge and human evaluations across all three models. 

\textbf{Lexical Similarity:} Traditional lexical similarity metrics yielded low absolute scores across all models and both evaluation methods. Metrics such as ROUGE-L, BLEU, and METEOR rely heavily on word overlap, which make them less suited for this task, but help in identifying data leakages and give an idea of word matches. Relatively, all the models performed similarly, although the proprietary models take a slight edge over DeepSeek v3.2. 

\textbf{Semantic Similarity:} The BERTScore provides a more accurate reflection of performance, measuring semantic embedding overlap rather than token matching. The consistent scores in the $0.55-0.60$ range for both ATAM and SAAM suggest that while the wording differed, the evaluation generated by the agentic systems was similar to the human baselines in structure and sense.

\textbf{LLM-as-a-Judge:} In the automated qualitative assessment, Claude 4.5 Sonnet notably outperformed other models in \textit{Relevance} and \textit{Completeness}. However, the results highlight a distinct divergence in model strengths. While Claude achieved superior scores in identifying comprehensive risks, Gemini 3 Flash consistently scored higher in \textit{Coherence} and \textit{Conciseness} across both ATAM and SAAM tasks. This is known as ``verbosity bias''~\cite{verbositybias}, where some models prefer to give verbose outputs even if the information in the response is diluted. This shows us a tradeoff in model performance when it comes to performing such complex tasks.

\subsection*{\textbf{RQ2.2} Human Evaluation Insights}

While automated metrics provide a baseline, the nuance of software architecture requires human judgment. Our human evaluation revealed that agentic systems are highly capable of replicating the \textit{form} of an evaluation, but differ in the \textit{substance} of their analysis. In this analysis, we concluded that even though all the models performed very closely, Claude 4.5 Sonnet produced the most complete software architecture evaluations, given our dataset as input.

A consistent trend observed across all models was the trade-off between coverage and focus. The agentic systems proved to be exhaustive, often identifying a higher volume of risks and scenarios than the human teams. As reflected in the \textit{Conciseness} scores (Table \ref{tab:results_master}), human evaluators penalized the models for listing low-impact risks alongside critical ones. The agents lacked the knowledge and experience required to filter out technically valid but business-irrelevant findings.

Despite the prioritization challenges, the agents demonstrated instances of reasoning that surpassed the source material. For example, in the ATAM evaluation for the \textit{Wargame 2000} system, Claude 4.5 Sonnet generated a logistical scenario regarding the time-cost analysis of realistic missile simulations, a critical concern that was absent in the original human-conducted study. This suggests that the agents can leverage their domain knowledge to uncover risks that human teams might overlook due to scope constraints.

\subsection*{\textbf{RQ3} Cost and Sustainability Impact}

To assess the practical viability of automating software architecture evaluation, we analyzed the computational and economic costs associated with each agentic workflow. Table \ref{tab:token_time_results} presents the average token consumption and execution time across our dataset.

\begin{table}[htbp]
\centering
\caption{Efficiency Metrics for Agentic Systems}
\label{tab:token_time_results}
\resizebox{0.5\textwidth}{!}{
\begin{tabular}{llccc}
\toprule
\textbf{Method} & \textbf{Model} & \textbf{Input Tokens} & \textbf{Output Tokens} & \textbf{Time Taken (s)} \\
\midrule
\multirow{3}{*}{\textbf{SAAM}} 
 & Gemini 3 Flash & 7129 & 2739 & 207 \\
 & DeepSeek v3.2 & 10282 & 4046 & 217 \\
 & Claude 4.5 Sonnet & 26138 & 17901 & 391 \\
\midrule
\multirow{3}{*}{\textbf{ATAM}} 
 & Gemini 3 Flash & 25730 & 7568 & 140 \\
 & DeepSeek v3.2 & 63167 & 7253 & 196 \\
 & Claude 4.5 Sonnet & 47517 & 17290 & 291 \\
\bottomrule
\end{tabular}
}
\end{table}

The data reveals a significant differentiation in efficiency driven by model architectures. Gemini 3 Flash demonstrated the most streamlined performance, requiring the lowest token count and execution time while simultaneously achieving the highest ratings in our human evaluation (refer to Table \ref{tab:results_master}). In contrast, Claude 4.5 Sonnet exhibited a high degree of verbosity, consuming nearly double the time and significantly more output tokens. This correlation suggests that while larger, more verbose models may offer detailed reasoning, they do not necessarily yield higher utility in structured evaluation tasks where conciseness is preferred.

Beyond computational metrics, the economic implications of this workflow are substantial. A traditional human-led ATAM is a resource-intensive process, typically requiring 3 to 4 person-weeks of total effort as stated by Clements et al. \cite{clements_2001}. Based on standard industry estimation techniques \cite{cost_estimation}, a single evaluation can cost between \$5,000 and \$10,000 when factoring in the billable hours of architects and managers. 

In comparison, our most expensive agentic run (Claude 4.5 Sonnet) incurred a compute cost of approximately \$0.40 and completed the task in under 5 minutes. While the agentic workflow is not yet a complete replacement for human expertise, this order-of-magnitude reduction in cost and time lowers the barrier to entry for architecture evaluation. It allows organizations to treat evaluation not as a costly, one-time ritual, but as a routine, continuous check, where human experts are brought in to review the high-priority risks identified by the agent.

\section{Discussion}
\label{sec:discussion}

Our results demonstrate that Agentic AI can effectively reduce the barriers to Software Architecture Evaluation. 

\subsection{Lessons Learned}

We observed that agents perform well at broad risk identification but struggle with relevance filtering. During the human evaluation, we noted that the outputs often contained too many scenarios. The agents frequently prioritized minor details, such as dependency mismatches, over actual structural concerns. This indicates that while agents can follow the basic steps of an evaluation, they lack the business intuition needed to find the most critical flaws. A potential solution for future work is context engineering. By providing the agents with stricter business guidelines and explicit project constraints in their initial prompts, we might be able to guide the models to filter out irrelevant details before the human review stage.

Furthermore, the quality of the output was heavily dependent on the description of the input. In the SAAM workflow, agents often defaulted to generic non-functional requirements when the input context was sparse. Conversely, the ATAM workflow demonstrated significantly higher quality when the \textit{Evaluation Inputs} included raw candidate scenarios extracted from the text. This suggests that agentic systems perform best when given specific stakeholder data, guiding them to the correct direction.

We also note that our inputs were structured summaries extracted from academic papers. In a real setting, architecture evaluation involves navigating missing information, conflicting stakeholder opinions, and large volumes of raw documentation. Because our agents received clean data, their performance likely represents a best case scenario. Applying this workflow in practice will require additional steps to process raw project documents before the core evaluation agents can begin their analysis.

\subsection{Implications for Researchers}




Capilla et al.~\cite{capilla10years} emphasized that widespread adoption of architecture knowledge management requires ``intelligent'' tools capable of semi-automating documentation and reasoning. This is essential to prevent ``knowledge vaporization'' in modern large scale software systems, where manually documenting is a major cause~\cite{5090042}. By demonstrating the potential of agents to act as active reasoning partners, we are moving closer towards the automation they envisioned.

To this end, the dataset curated in this study provides a starting point for validating specialized models for software architectural evaluation tasks. This dataset can be used to further create and evaluate such systems, validating their real-world applicability. Additionally, our observation that agents occasionally identified valid risks missed by human evaluators suggests a promising avenue for research into hybrid workflows, where the human is more involved in the agentic workflow, acting as a partner rather than basic enablers. 

\subsection{Implications for Practitioners}

Our preliminary results suggest that agentic systems are best used as drafting tools rather than independent evaluators. Babar et al.~\cite{sae_adoption_barriers} note that time and resource costs prevent many organizations from adopting evaluation methods. Our workflow helps lower this barrier by handling the initial phase of data collection and scenario generation.

In practice, a human architect would not accept the agent output as a final report. Instead, they would use the tool as a starting point. The human user would first provide the available architecture documents to the system. The agents then process this data to generate a draft containing potential scenarios and risks. At this point, the human architect must step in to review the generated JSON file. They will manually delete scenarios that are technically valid but practically unimportant for the business, and they will adjust the priority levels of the remaining items. 

The most important human task in this workflow is writing the final recommendations. While the agents can list structural problems, preparing a convincing summary for management requires human experience. Proposing a realistic architectural change requires an understanding of company budgets and team capabilities, which the agents do not have. Therefore, the tool acts as a brainstorming assistant that helps reduce the initial blank page problem, rather than an autonomous decision maker. The human architect remains responsible for filtering the data and making the final decisions.



\section{Threats to Validity}
\label{sec:threats_to_validity}

We follow the distinction outlined by Wohlin et al.~\cite{wohlin} and discuss the efforts taken to mitigate identified threats to the validity of our study.

\subsection{External Validity}
The primary threat lies in the selection of specific LLMs and the design of the agent personas. Due to limitations with number of LLMs we can experiment with, the training data of selected LLMs might produce a bias. To mitigate this, we selected models based on the LMSYS Chatbot Arena Leaderboard to ensure they represent the state-of-the-art at the time of conducting the study. Furthermore, the design of our agents is strictly based on the stakeholder definitions and protocols defined by Kazman et al. in the original ATAM and SAAM studies~\cite{kazman_atam,kazman_saam} to ensure the workflow remains methodologically sound.

\subsection{Internal Validity}
Internal validity threats arise from the configuration of the models and the inherent bias in automated evaluation metrics. We addressed this by maintaining identical hyperparameter configurations across all experimental runs. While the use of an LLM-as-a-Judge introduces self-preference bias~\cite{llm_judge_bias}, we mitigated this by contrasting these scores with manual human evaluation. Similarly, while a single model was utilized for the initial data extraction, the data was validated using manual author validation.

\subsection{Construct Validity}
The subjective nature of software architecture evaluation poses a challenge to construct validity. To minimize interpretation bias regarding the quality of the generated reports, the human evaluation was conducted independently by two authors. This separation ensured no personal bias influenced the scores. Additionally, the creation of the dataset involved a dual-verification process where both authors explicitly validated each extracted data point, ensuring that the ground truth represented the findings of the original case studies.

Another major threat to construct validity is the nature of our input data. Real software architecture evaluation involves navigating messy, incomplete, and sometimes contradictory documentation. It also involves speaking directly with stakeholders to uncover hidden requirements. In this study, the agents evaluate academic papers and case study summaries. These documents are already condensed and structured by human authors. In some cases within our dataset, the provided architecture descriptions are brief and lack the technical depth found in industrial settings. 

As a result, the agents operate on simplified, pre-processed information rather than raw enterprise artifacts. Because they only know what is in the text, they cannot ask follow up questions to real users to fill in missing gaps. This limits the generalizability of our results. This study primarily tests if agents can follow the structured steps of ATAM and SAAM when given clean data. It does not prove that the system can currently handle the noise and complexity of a real world corporate environment. Future work must test this multi-agent approach on raw, unstructured enterprise data to better understand its practical limitations and fail cases.

\section{Conclusion and Future Work}
\label{sec:conclusion}

This paper investigated the feasibility of automating software architecture evaluation through the use of agentic AI workflows. By establishing a validated dataset and designing role-based agents, we demonstrated that LLMs in an agentic setup can effectively replicate the procedural steps of ATAM and SAAM. Our findings indicate that while these agents significantly reduce the time and cost associated with evaluation, they currently function best as helpers rather than replacements. They excel at broad risk identification and scenario generation, enabling organizations to transition from sporadic, high-cost reviews to continuous architectural assessment, provided that human expertise remains in the loop.

Future research should focus on expanding the open dataset to include a wider variety of architectural styles and domain-specific case studies. It can also be extended to other forms of software architecture evaluation methods, since it currently only supports SAAM and ATAM. Another possible direction is to evolve the current workflow into a fully dynamic, conversational system. Specifically, turning the \textit{Customer} agent into an actual human who can provide any necessary information asked by the Evaluation Team. This will more closely simulate the conversational nature of such architectural evaluations.

\section{Data Availability}
\label{sec:data_availability}

The dataset consisting of 39 structured architectural evaluation case studies and the source code for the agentic workflow, including outputs and final results, are publicly available. These resources can be accessed via our replication package at \url{https://zenodo.org/records/21374126}.

\bibliographystyle{ACM-Reference-Format}
\bibliography{sample-base}

\end{document}